\documentclass[twocolumn]{jpsj2} 

\def\runtitle{Instructions for the Preparation of Manuscript}
\def\runauthor{Online-Journal Subcommittee}

\title{
In-plane Anisotropy of the Magnetic Fluctuations in 
Na$_{x}$CoO$_{2} \cdot$\textit{y}H$_{2}$O
}
\author{
C. Michioka, Y. Itoh, H. Ohta, M. Kato$^1$ and K. Yoshimura
}

\inst{
Department of Chemistry, Graduate School of Science,\\
Kyoto University, Kyoto 606-8502, Japan \\ 
$^1$Department of Molecular Science and Technology, Faculty of Engineering,\\
Doshisha University, Kyotanabe, Kyoto 610-0394, Japan
}

\recdate{\today}

\abst{
We report the $^{59}$Co NMR studies of the in-plane anisotropy of bilayer hydrated Na$_{x}\mathrm{CoO_{2}} \cdot$$y\mathrm{H_{2}O}$ using a oriented powder sample by a magnetic field in Fluorinert FC70. We found for the first time the $ab$-plane anisotropy of the $^{59}$Co NMR Knight shift $K$, the nuclear spin-lattice relaxation rate 1/$T_{1}$ and the nuclear spin-spin relaxation rate 1/$T_{2}$ at a magnetic field $H \sim$ 7.5 T up to 200K. Below 75 K, the anisotropy of $K$ is large compared with that at high temperatures. The hyperfine coupling constants seem to change around the temperature 150 K, in which the bulk susceptibility $\chi$ shows broad minimum, suggesting a change of the electronic state of CoO$_{2}$ plane. 1/$T_{1}$ also shows a significant anisotropy, which cannot be explained only by the anisotropy of the hyperfine coupling constants nor the anisotropic uniform spin susceptibility. The difference in the in-plane anisotropy of $T_{1}$ from that of $K$ indicates that the magnetic fluctuation at a finite wave vector $\vec{q} \neq 0$ is also anisotropic and the anisotropy is different from that at $\vec{q} = 0$. 
}

\kword{
superconductivity, sodium cobalt oxide, NMR, anisotropy
}

\begin{document}

\maketitle
Since the discovery of the superconductivity on the triangular CoO$_{2}$ plane in Na$_{x}\mathrm{CoO_{2}}\cdot$$y\mathrm{H_{2}O}$ \cite{Takada}, not only the superconducting state but also the normal state have been intensively studied to understand the unique properties and the electronic structure. In contrast to the high-$T\mathrm{_{c}}$ cuprate, in the case of cobaltate it is necessary to consider a $t_{2g}$-orbital magnetism of Co$^{4+}$ ($d^{5}$). Sodium deficiencies and oxonium ions introduce a mixing state of $S$ = 0 Co$^{3+}$ and $S$ = 1/2 Co$^{4+}$. The $\mathrm{CoO_{2}}$ layers in Na$_{x}\mathrm{CoO_{2}}\cdot$$y\mathrm{H_{2}O}$ consist of edge shared CoO$_{6}$ octahedra which are compressed along the trigonal axis. The trigonal distortion and spin-orbit interaction divide the $t_{2g}$ level into three levels (the $f$, $g$, and $h$ states labeled in Ref. [2]). In Co$^{4+}$, a hole with pseudospin 1/2 resides on the $f$ level, whose wave function contains both $e'_{g}$ ($|l_{z} = \pm1\rangle$) and $a_{1g}$ ($|l_{z} = 0\rangle$) states mixed up by a spin-orbit coupling \cite{Khaliullin}. Although the orbital moment can give an anisotropy to the magnetic moment even in the itinerant system, the microscopic and macroscopic details of in-plane magnetic anisotropy in Na$_{x}\mathrm{CoO_{2}}\cdot$$y\mathrm{H_{2}O}$ remain to be explored so far. 

Due to the nature of the soft chemical treatment, it is difficult to intercalate the water molecules homogeneously into Na$_{x}$CoO$_{2}$ to prepare a bilayer hydrated sample, especially in a large single crystal. Therefore, in the present stage, the bilayer hydrated powder samples are still more homogeneous than single crystals and reliable to study the intrinsic physical properties. In the previous work, we observed the two-dimensional (2D) powder pattern of $^{59}$Co NMR by using an organic solvent hexane to fix the powders under a magnetic field of 8 T \cite{Kato}. However, hexane is not convenient due to a volatile property and then we could not obtain the higher temperature data. We need a convenient solvent, which does not absorb water molecules and have an appropriate melting point. Recently, we found that Fluorinert is most suitable for fixing powders of Na$_{x}\mathrm{CoO_{2}} \cdot$$y\mathrm{H_{2}O}$ and we succeeded in measuring the well-oriented 2D powder data up to 200K. 

In this paper, we report for the first time the $ab$-plane anisotropy of the $^{59}$Co NMR Knight shift $K$, the nuclear spin-lattice relaxation rate 1/$T_{1}$ and the nuclear spin-spin relaxation rate 1/$T_{2}$ at a magnetic field $H \sim$ 7.5 T up to 200 K for the bilayer hydrated Na$_{x}\mathrm{CoO_{2}} \cdot$$y\mathrm{H_{2}O}$ with $T\mathrm{_{c}}$ = 4.8 K. We found that $K_{x}$ strongly depends on temperature while $K_{y}$ increases slightly with the decrease of temperature. We also obtained the striking result that 1/$T_{1}$ at $H\parallel x$-axis is about two times larger than that at $H\parallel y$-axis. The $T_{1}$ anisotropy cannot be explained only by the anisotropic hyperfine coupling constants. 

\begin{figure}[t]
\begin{center}
\includegraphics[width=0.8\linewidth]{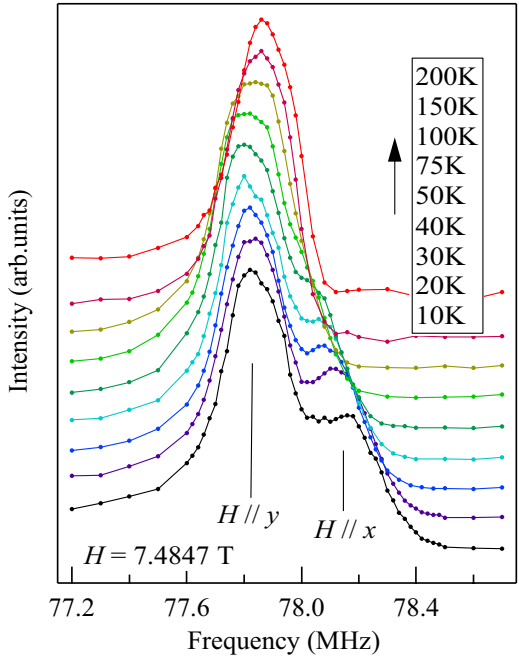}
\end{center}
\caption{\label{fig:fig1}
Center lines ($I_z = - 1/2 \leftrightarrow + 1/2$) of the frequency-swept $^{59}$Co NMR spectra in Na$_{x}\mathrm{CoO_{2}} \cdot$$y\mathrm{H_{2}O}$ under a magnetic field of 7.4847 T. Two-peak structure is explained by the anisotropic Knight shifts but not the asymmetry parameter $\eta$.
}
\end{figure}

The $^{59}$Co NMR and nuclear quadrupole resonance (NQR) were carried out for the optimal $T\mathrm{_{c}}$ = 4.8 K sample of bilayer hydrated Na$_{x}\mathrm{CoO_{2}} \cdot$$y\mathrm{H_{2}O}$, which gives a optimal superconductivity in the phase diagram \cite{Ihara,Michioka}. The powders of Na$_{x}\mathrm{CoO_{2}} \cdot$$y\mathrm{H_{2}O}$ were oriented under a magnetic field of $H \sim$ 7.5 T by Fluorinert FC70 (melting point of 248 K). In this situation, the $c$-axis is oriented perpendicular to the external field and the $c$-plane is oriented randomly along the magnetic field. All the frequency-swept $^{59}$Co NMR spectra at the center resonance ($I_z = - 1/2 \leftrightarrow + 1/2$) were measured under a magnetic field of 7.4847 T with a conventional spin-echo technique. The $^{59}$Co nuclear spin-lattice relaxation time $^{59}T_{1}$ was measured by an inversion recovery technique. The spin-echo signal $M(t)$ was measured as a function of long delay time $t$ after an inversion pulse, and $M(\infty)[\equiv M(t>10T_1)]$ was recorded. The nuclear spin-spin relaxation time $^{59}T_{2}$ was measured by the spin-echo decay.

Figure \ref{fig:fig1} shows the 2D powder $^{59}$Co NMR spectra between 10 and 200 K. These resonance lines arise from the transition, $I_z = \pm 5/2 \leftrightarrow \pm 7/2$ of $^{59}$Co (nuclear spin $I$ = 7/2). In all data, the spin-echo intensities at the frequencies far from the resonance center are close to zero, suggesting in-plane well-oriented 2D powder spectra. The spectrum at 10 K shows a two-peak structure, which is explained by a lower symmetric spectrum than a uniaxial symmetry \cite{Kato}. The peaks at lower and higher frequencies correspond to $H\parallel y$ and $H\parallel x$, respectively. Although the effect of a non-uniaxial electric field gradient with a finite asymmetry parameter $\eta$ would split the resonance peak, the large splitting of the observed spectrum cannot be explained by $\eta$ which is estimated by the NQR measurement (see below). The temperature dependence of the peak frequencies is mainly due to the temperature dependence of the Knight shifts but not the quadrupole shift. The peak frequency of $K_{y}$ slightly decreases with increasing temperature, while that of $K_{x}$ decreases markedly. The $K_{y}$ and $K_{x}$ peaks can be clearly distinguishable below 75 K. At 100 and 150 K, the $K_{y}$ and $K_{x}$ peaks are not clearly split, suggesting that $K_{y}$ is similar to $K_{x}$. At 200 K, the spectrum can be fitted by a double Gaussian function.

\begin{figure}[t]
\begin{center}
\includegraphics[width=0.85\linewidth]{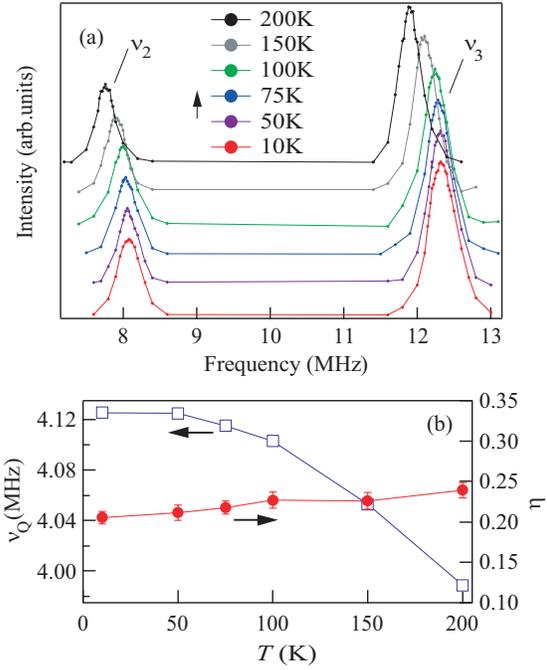}
\end{center}
\caption{\label{fig:fig2}
(a) Temperature dependence of $^{59}$Co NQR spectra. Lower and higher frequency peaks correspond to $\nu_{2}$ ($I_z = \pm 3/2 \leftrightarrow \pm 5/2$) and $\nu_{3}$ ($I_z = \pm 5/2 \leftrightarrow \pm 7/2$), respectively. (b) Temperature dependence of $\nu\mathrm{_{Q}}$ and $\eta$ are calculated from the $\nu_{2}$ and $\nu_{3}$.
}
\end{figure}

In order to estimate quantitatively the quadrupole shift and decide the intrinsic Knight shift, we carried out the $^{59}$Co NQR measurements. Figure \ref{fig:fig2}(a) shows the $^{59}$Co NQR spectra between 10 and 200 K. The lower and higher frequency peaks in the NQR spectra correspond to the resonances $I_z = \pm 3/2 \leftrightarrow \pm 5/2$, $\nu_{2}$ and $I_z = \pm 5/2 \leftrightarrow \pm 7/2$, $\nu_{3}$, respectively. We estimated a quadrupole resonance frequency $\nu\mathrm{_{Q}}$ and an asymmetric parameter $\eta \equiv (\partial^{2}V/\partial x^{2}-\partial^{2}V/\partial y^{2})/(\partial^{2}V/\partial z^{2})$ from $\nu_{2}$ and $\nu_{3}$ \cite{Abragam}. The nuclear spin Hamiltonian of the quadrupole interaction is expressed as
\begin{align}
H_{Q} = a[3I_{z}^{2}-I(I+1)+\frac{1}{2}\eta(I_{+}^{2}+I_{-}^{2})],
\end{align}
\begin{align}
a = \frac{e^{2}qQ}{4I(2I-1)} = \frac{\nu_{Q}h}{6}.
\end{align}
Then the eigen value equation for $I$ = 7/2 is written as
\begin{align}
E^{4}-42(1+\frac{1}{3}\eta^{2})(3a)^{2}E^{2}-64(1-\eta^{2})(3a)^{3}E \notag\\
+105(1+\frac{1}{3}\eta^{2})^{2}(3a)^{4} = 0.
\end{align}
The eigen values of four Kramers doublets ($E_{\pm7/2}$, $E_{\pm5/2}$, $E_{\pm3/2}$, $E_{\pm1/2}$) were numerically solved. Then we obtained $\nu_{3} \equiv E_{\pm7/2} - E_{\pm5/2}$ and $\nu_{2} \equiv E_{\pm5/2} - E_{\pm3/2}$ as functions of $\nu\mathrm{_{Q}}$ and $\eta$. 
$\eta$ was determined from the ratio of $\nu_{2}/\nu_{3}$, and then the quaqrapole frequency $\nu\mathrm{_{Q}}$ was estimated from $\eta$ and $\nu_{3}$. The obtained $\nu\mathrm{_{Q}}$ and $\eta$ are shown in Fig. \ref{fig:fig2}(b).
$\nu\mathrm{_{Q}}$ is almost invariant between 10 and 50 K and decreases as temperature increases up to 200 K, while $\eta$ increases slightly.

\begin{figure}[t]
\begin{center}
\includegraphics[width=0.85\linewidth]{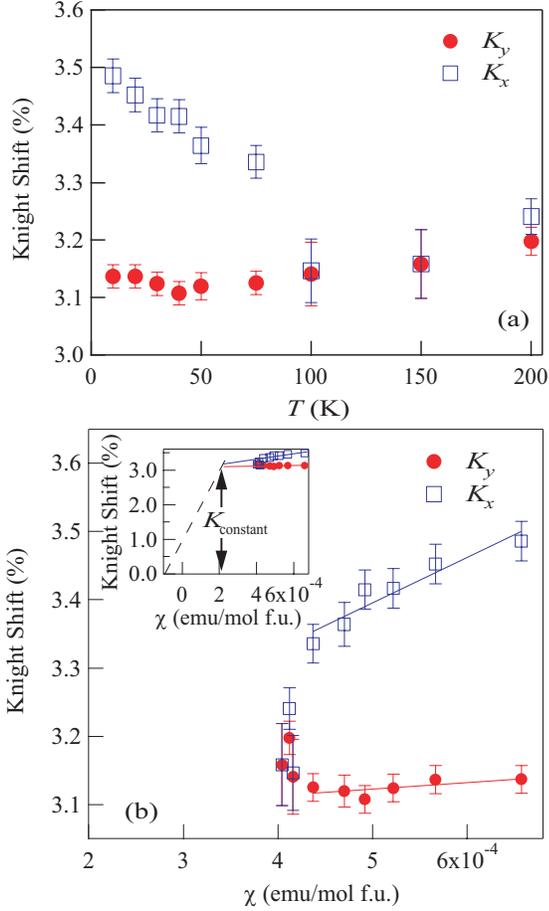}
\end{center}
\caption{\label{fig:fig3}
(a) $^{59}$Co Knight shifts $K_{x}$ and $K_{y}$ of Na$_{x}\mathrm{CoO_{2}} \cdot$$y\mathrm{H_{2}O}$. The shifts were estimated with taking account of the quadrupole shift. (b) $K$-$\chi$ plot. The bulk susceptibility $\chi$ of powder sample was used as a horizontal axis.
}
\end{figure}

Figure \ref{fig:fig3}(a) shows the temperature dependence of the Knight shifts determined by using the NQR parameters and second-order perturbation theory \cite{Kato}. As mentioned above, the two peaks of the spectra are mainly due to the in-plane anisotropy of the Knight shift but not that of the quadrupole shift. From the analysis of the shape of NMR spectra, at 100 and 150 K the Knight shifts are almost isotropic in the $ab$-plane and once again slightly anisotropic at 200 K. Below 100K, the distinguishable two-peak structure of NMR spectra can be explained by the $ab$-plane anisotropic Knight shifts $K_{x}$ and $K_{y}$. $K_{x}$ increases monotonically with decreasing temperature, while $K_{y}$ slightly increases. Thus, it turned out that the low temperature Curie-Weiss-type upturn of the magnetic susceptibility \cite{Kato} is intrinsic.

Figure \ref{fig:fig3}(b) is the $K$-$\chi$ plot, where temperature is an implicit parameter. In this plot, we tentatively employ the bulk magnetic susceptibility $\chi$ of powder sample. We believe that $\chi$ is anisotropic, especially in $ab$-plane at low temperatures. Below the room temperature, the bulk susceptibility $\chi$ decreases with the decrease of temperature and shows a broad minimum around 150 K and again increases down to 10 K. Since the two peaks corresponding to $H\parallel x$ and $H\parallel y$ are obviously distinguishable and the two distinct lines are held below 75 K, we can estimate the hyperfine coupling constants as $A_{x}$ = +37.2 kOe/$\mu\mathrm{_{B}}$ and $A_{y}$ = +5.2 kOe/$\mu\mathrm{_{B}}$, respectively, from the data below 75 K. The value of $A_{x}$ is similar to that previously estimated while $A_{y}$ is less than half \cite{Kato}. The differences between the present and the previous \cite{Kato} works are mainly due to the imperfect orientation by a magnetic field in the previous work. 

Above 100 K, $K_{x}$ is close to $K_{y}$. The linear relations in the $K$-$\chi$ plot do not seem to hold above 100 K. Thus, we can conclude that the hyperfine coupling constants at high temperatures are different from that at low temperatures. Then, the electronic state at the cobalt site is thought to be changed at about the temperature $\sim$ 150 K, where the bulk $\chi$ shows the minimum behavior. The change may be attributed in a motion of sodium ions or a tendency of superlattice formation. The anomalies found in 1/$T_{1}$ of the $^{23}$Na NMR implies a sodium contribution to the change of the electronic state in the CoO$_{2}$ plane \cite{Ohta2}. As shown in the inset figure, a temperature independent Knight shift, which consists of contributions of the diamagnetism, the orbital part and maybe a temperature-independent spin part, is estimated about 3 $\%$. The detailed analysis has already been reported \cite{Kato}. 

\begin{figure}[t]
\begin{center}
\includegraphics[width=0.9\linewidth]{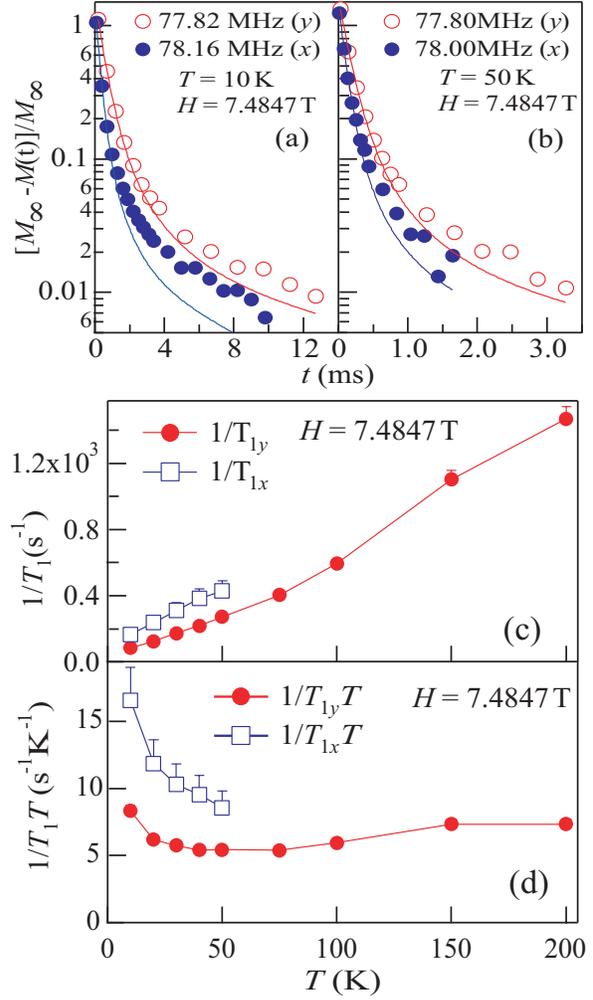}
\end{center}
\caption{\label{fig:fig4}
Recovery curves at (a) 10 and (b) 50 K. Open and closed circles represent the data measured at the peak frequencies of $K_{y}$ and $K_{x}$, respectively. Solid lines are fitted lines by the theoretical relaxation function. Temperature dependence of  (c) 1/$T_{1}$ and (d) 1/$T_{1}T$.
}
\end{figure}

Figures \ref{fig:fig4}(a) and (b) show the $^{59}$Co nuclear spin-echo recovery curves at $H\parallel y$-axis (lower frequency) and at $H\parallel x$-axis (higher frequency) at 10 and 50 K, respectively. The recovery curves were fitted by the theoretical relaxation function of the central resonance, as follows: 
\begin{align}
\frac{M(\infty)-M(t)}{M(\infty)} \notag\\
= M_{0}\left( \frac{1}{84}e^{\frac{-t}{T_1}} + \frac{3}{44}e^{\frac{-6t}{T_1}} + 
\frac{75}{364}e^{\frac{-15t}{T_1}} + \frac{1225}{1716}e^{\frac{-28t}{T_1}}
\right),
\label{eq:recovery}
\end{align}
where $M_{0}$ and $T_{1}$ are fitting parameters. The obtained recovery curves at the peak-frequency of $K_{y}$ agree with theoretical curves. However, a small slow component at $K_{x}$ exists maybe due to a small contribution of $K_{y}$, and then the experimental curves were fitted by the theoretical curves up to 1/10 decay. One can find obvious differences in recovery curves at $H\parallel y$ and at $H\parallel x$. 

Figure \ref{fig:fig4}(c) shows the temperature dependence of the nuclear spin-lattice relaxation rate at the peak frequencies of $H\parallel y$, 1/$T_{1y}$ and that of $H\parallel x$, 1/$T_{1x}$. Figure \ref{fig:fig4}(d) shows 1/$T_{1i}T$ ($i = x, y$). The temperature dependence of 1/$T_{1y}$ is similar to that at $H$ = 0 T, which was measured by NQR \cite{Kato}, however, the absolute values are smaller, suggesting the suppression of the magnetic fluctuations by the magnetic field. 1/$T_{1x}$ is clearly different from 1/$T_{1y}$ below 50 K. 

In Fig. \ref{fig:fig4}(d), we obtained the significant result that 1/$T_{1x}T$ is about two times larger than 1/$T_{1y}T$. Since 1/$T_{1i}$ ($i = x, y$) should arise from the magnetic fluctuations $(\delta h)^{2}$ perpendicular to the $i$ direction, 1/$T_{1x}$ and 1/$T_{1y}$ should be proportional to $(\delta h_{z})^{2}+(\delta h_{y})^{2}$ and $(\delta h_{z})^{2}+(\delta h_{x})^{2}$, respectively \cite{Moriya}. $(\delta h_{i})^{2}$ ($i = x, y, z$) is the square of local magnetic field fluctuations along the $i$-axis. Then the Knight shifts should be $K_{x}^{s} = A_{x} \chi_{x}$ and $K_{y}^{s} = A_{y} \chi_{y}$. 1/$T_{1x}$ is about two times larger than 1/$T_{1y}$, while $K_{y}$ is smaller than $K_{x}$. Therefore, the anisotropy of $T_{1}$ cannot be explained only by the anisotropy of the hyperfine coupling constants nor the anisotropic uniform spin susceptibility. The difference in the in-plane anisotropy of $T_{1}$ from that of $K$ indicates that the magnetic fluctuation at a finite wave vector $\vec{q} \neq 0$ is also anisotropic and the anisotropy is different from that at $\vec{q} = 0$. 

\begin{figure}[t]
\begin{center}
\includegraphics[width=0.92\linewidth]{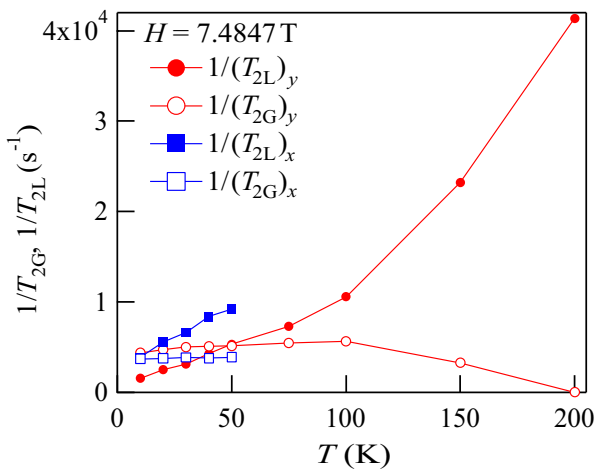}
\end{center}
\caption{\label{fig:fig5}
Temperature dependence of 1/($T\mathrm{_{2G}}$)$_{i}$ and 1/($T\mathrm{_{2L}}$)$_{i}$ ($i = x, y$). 
}
\end{figure}

Figure \ref{fig:fig5} shows the temperature dependence of the transverse relaxation rates of a Gaussian component 1/($T\mathrm{_{2G}}$)$_{i}$ and a Lorentzian component 1/($T\mathrm{_{2L}}$)$_{i}$ ($i = x, y$). Behaviors of 1/($T\mathrm{_{2G}}$)$_{i}$ and 1/($T\mathrm{_{2L}}$)$_{i}$ are similar to that of the nonsuperconducting bilayer hydrated Na$_{x}\mathrm{CoO_{2}} \cdot$$y\mathrm{H_{2}O}$ in NQR measurement \cite{Michioka}. As well as 1/$T_{1}$, especially 1/($T\mathrm{_{2L}}$)$_{i}$ shows a strong anisotropy for $x$ and $y$. Therefore, the primary contribution to 1/($T\mathrm{_{2L}}$)$_{i}$ is a $T_{1}$ process and 1/($T\mathrm{_{2G}}$)$_{i}$ is nuclear spin-spin relaxation rate. At high temperatures, the 1/($T\mathrm{_{2G}}$)$_{i}$ contribution becomes quite small and a part of the transverse relaxation may be originated in a contribution of the sodium motion.

The magnetic susceptibility measured on the single crystal was found to be anisotropic in $H$ parallel and perpendicular to the CoO$_{2}$ plane \cite{Chou}. However, the in-plane anisotropy is not reported. From the temperature-dependent anisotropy of $K$ and 1/$T_{1}$ of the $^{59}$Co NMR as well as the anomalous behaviors of the $^{23}$Na NMR \cite{Ohta2} and the electrical resistivity \cite{Jin}, one may conclude that the electronic state of CoO$_{2}$ plane is different at low and high temperatures, which may be due to the change of the charge-transfer state from sodium and oxonium ions. Unquenched orbital moments may play a significant role in the low temperature state. Unconventional electronic states such as hidden Kagom\'{e} symmetry in the CoO$_{2}$ plane \cite{Koshibae} were proposed from the study of the $t_{2g}$ degeneracy. A variety of the superconducting properties was also studied by taking account of the spin-orbit coupling \cite{Khaliullin,Yanase}. The presence of the $ab$-plane anisotropy of the magnetic fluctuations is a key to reveal the superconducting mechanism in Na$_{x}\mathrm{CoO_{2}} \cdot$$y\mathrm{H_{2}O}$.

This work is supported by Grant-in Aid for Scientific Research on Priority Area "Invention of anomalous quantum materials", from the Ministry of Education, Culture, Sports, Science and Technology of Japan (Grant No. 16076210).

\end{document}